\documentclass[aps,prd,amsmath, amssymb,eqsecnum,
nofootinbib,showpacs,tightenlines]{revtex4}
\global\arraycolsep=2pt
\usepackage{bm}
\usepackage{epsfig}

\newcommand{\ben}{\begin{equation*}}
\newcommand{\een}{\end{equation*}}
\newcommand{\bean}{\begin{eqnarray*}}
\newcommand{\eean}{\end{eqnarray*}}

\newcommand{\nn}{\nonumber}
\newcommand{\be}{\begin{equation}}
\newcommand{\ee}{\end{equation}}
\newcommand{\bea}{\begin{eqnarray}}
\newcommand{\eea}{\end{eqnarray}}

\begin{document}
\title{Exact Casimir energies at nonzero temperature: Validity of
proximity force approximation and interaction of semitransparent
spheres}

\author{Kimball A. Milton} 
\email{milton@nhn.ou.edu}
\homepage{http://www.nhn.ou.edu/

\author{Prachi Parashar}
\email{prachi@nhn.ou.edu}

\author{Jef Wagner}
\email{wagner@nhn.ou.edu}
\affiliation{Oklahoma Center for High Energy Physics 
and Homer L. Dodge Department of Physics and Astronomy,
University of Oklahoma, Norman, OK 73019-2061, USA}

\author{K.V. Shajesh}
\email{shajesh@nhn.ou.edu}
\affiliation{St.~Edwards School, Vero Beach, Florida, 32963-2699,
USA}
\date{\today}

\begin{abstract}
In this paper, dedicated to the career of Tom Erber, we consider the 
Casimir interaction between weakly coupled bodies at nonzero temperature.  
For the case of semitransparent bodies, that is, ones described by 
$\delta$-function potentials, we first
examine the interaction between an infinite plane and an arbitrary
curved surface.  In weak coupling, such an interaction energy coincides
with the exact form of the proximity force approximation obtained by summing
the interaction between opposite surface elements at arbitrary temperature.  
This result generalizes a theorem proved recently by Decca et al.
We also obtain exact closed-form
 results for the Casimir energy at arbitrary temperature
for weakly-coupled semitransparent spheres.
\end{abstract}

\maketitle

\section{Introduction}

Since the earliest calculations of fluctuation forces between bodies
\cite{Casimir:1948dh},
that is, Casimir or quantum vacuum forces, multiple scattering methods
have been employed.  Rather belatedly, it has been realized that
such methods could be used to obtain accurate numerical results
in many cases \cite{Wirzba:2007bv, Bordag:2008gj, maianeto08, Emig:2007cf}. 
These results allow us to transcend the limitations of the
proximity force theorem (PFT) \cite{pft,derjaguin}, and so make better
comparison with experiment, which typically involve curved
surfaces. (For a review of the experimental
situation, see Ref.~\cite{Onofrio:2006mq}.)

The multiple scattering formalism, which is in principle exact,
dates back at least into the 1950s \cite{krein,renne}.  Particularly
noteworthy is the seminal work of Balian and Duplantier \cite{balian}.
(For more complete references see Ref.~\cite{Milton:2007wz}.)
This technique, which has been brought to a high state of perfection
by Emig et al.\ \cite{Emig:2007cf}, 
has concentrated
on numerical results for the Casimir forces between conducting and
dielectric bodies such as spheres and cylinders. For recent
impressive numerical results for metals and dielectrics
see Refs.~\cite{rahi,reid}. Our group has
noticed that the multiple-scattering method can yield exact, closed-form 
results for bodies that are weakly coupled to the quantum field
\cite{Milton:2007gy,Milton:2007wz}. 
(That is, we are carrying out first-order perturbation
theory in the background potential.  For early examples of this in the
Casimir context, see Ref.~\cite{bordag90}.)
This allows an exact assessment
of the range of applicability of the PFT.  The calculations there,
however, as those in recent extensions of our methodology
\cite{CaveroPelaez:2008tj}, 
have been restricted to scalar fields with $\delta$-function potentials,
so-called semitransparent bodies.  
(These are closely related to plasma shell models \cite{Bordag:2008gj,
Bordag:2008rc,Bordag:2007zz,barton,Bordag:2006kx}.) The
technique was recently extended to dielectric bodies \cite{Milton:2008vr,
Milton:2008bb}, characterized by a permittivity $\varepsilon$.
Strong coupling would mean a perfect metal, $\varepsilon\to\infty$, while
weak coupling means that $\varepsilon$ is close to unity.  

In this paper we will extend the weak-coupling formalism to the
situation of nonzero temperature.  
This extension is extremely straightforward.  We then apply the
general formula to the case of an arbitrarily curved semitransparent
surface above an infinite semitransparent plane.  Remarkably, the
result coincides with the use of the so-called proximity force
approximation (PFA), which in its general form is exact in this case
for all separations between the surfaces and for all temperatures.
We also obtain exact closed-form
 results for the forces between separated spherical
shells for all temperatures. In the Appendix we discuss exact
formulas for arbitrary positive and negative powers of the distances
between points on two spheres, needed for such calculations.
\section{Multiple scattering derivation of vacuum energy between 
weakly coupled potentials}

The quantum vacuum energy for the interaction mediated by
a massless scalar field between two 
nonoverlapping potentials $V_1(\mathbf{x})$ and $V_2(\mathbf{x})$ is
\be
E=-\frac{i}2\mbox{Tr}\, \ln(1-V_1G_1V_2G_2),\label{trln}
\ee
in terms of the single potential Green's functions
\be
G_i=(1+G_0V_i)^{-1}G_0.
\ee
The free Green's function, satisfying
\be
-\partial^2 G_0(x-x')=\delta(x-x'),
\ee
has the explicit form
\be
G_0(x-x')=\int\frac{d\omega}{2\pi}\mathcal{G}_0(\mathbf{r-r'},\omega) 
e^{-i\omega(t-t')},
\ee
where the time-Fourier transform is
\be
\mathcal{G}_0(\mathbf{r-r'},i\zeta)=\frac{e^{-|\zeta||\mathbf{r-r'}|}}
{4\pi|\mathbf{r-r'}|},
\ee
where we have performed the Euclidean rotation $\omega\to i\zeta$.

For weak potentials, the energy (\ref{trln}) simplifies dramatically:
\be
E\approx\frac{i}2\mbox{Tr}\,V_1 G_0V_2 G_0
=-\frac1{64\pi^3}\int (d\mathbf{r})(d\mathbf{r'})\frac{V_1(\mathbf{r})
V_2(\mathbf{r'})}{|\mathbf{r-r'}|^3}.\label{0T}
\ee

At finite temperature the integral over imaginary frequency becomes the
Matsubara sum:
\be
\int_{-\infty}^\infty \frac{d\zeta}{2\pi}\frac{e^{-2|\zeta||\mathbf{r-r'}|}}
{|\mathbf{r-r'}|^2}\to T\sum_{m=-\infty}^\infty \frac{e^{-4\pi T |m|
|\mathbf{r-r'}|}}{|\mathbf{r-r'}|^2}
=\frac{T}{|\mathbf{r-r'}|^2}\coth 2\pi T|\mathbf{r-r'}|,
\ee
so the interaction energy becomes
\be
E_T=-\frac{T}{32\pi^2}\int(d\mathbf{r})(d\mathbf{r'})V_1(\mathbf{r})
V_2(\mathbf{r'})\frac{\coth 2\pi T|\mathbf{r-r'}|}{|\mathbf{r-r'}|^2},
\label{ET}
\ee
which evidently reduces to Eq.~(\ref{0T}) for $T=0$.

\section{Parallel plates}
For parallel, semitransparent plates, separated by a distance $a$,
 where the potentials are
\be
V_1(\mathbf{r})=\lambda_1\delta(z),\quad V_2(\mathbf{r})=\lambda_2\delta(z-a),
\ee
the integrals in Eq.~(\ref{0T}) are readily carried out, with the resulting
energy per unit area $A$, $\mathcal{E}=E/A$:
\be
\mathcal{E}=-\frac{\lambda_1\lambda_2}{32\pi^2 a}.
\ee
This well-known result holds even if {\it one\/} of the plates has a finite
area $A$.  At finite temperature the result is
\be
\mathcal{E}_T=-\frac{\lambda_1\lambda_2 T}{16\pi}\int_{2\pi T a}^\infty
\frac{dx}x \coth x.\label{ppe}
\ee
The energy is ambiguous because it depends on the arbitrarily chosen upper
limit. However, it corresponds to a well-defined pressure between the plates,
\be
P_T=-\frac\partial{\partial a}\mathcal{E}_T=-\frac{\lambda_1\lambda_2 T}
{16\pi a}\coth 2\pi T a.
\ee

\section{Interaction between an infinite plane and an arbitrarily
curved surface: PFA}
Now consider the interaction between a semitransparent plane, described
by the potential
\be
V_1(\mathbf{r})=\lambda_1\delta(z),
\ee
and an arbitrary curved surface $S$, which does not intersect the plane $z=0$,
which corresponds to the potential
\be
V_2(\mathbf{r})=\lambda_2\delta(z-s(x,y)),
\ee
where $z=s(x,y)$ is the equation of the surface.
Then, from Eq.~(\ref{ET}) it is immediate that the energy 
is (the upper limit of the $x$ integration is again physically irrelevant)
\be
E_T=-\frac{\lambda_1\lambda_2 T}{16\pi}\int dS\int_{2\pi T z(S)} dx
\frac{\coth x}x,
\ee
where the area integral is over the curved surface.
This is precisely what one means by the proximity force approximation,
where one sums energies between adjacent elements treated as parallel
plates:
\be
E_{\rm PFA}=\int dS\mathcal{E}_\|(z(S)),
\ee
in view of Eq.~(\ref{ppe}).  This is in fact just the theorem proved
by Decca et al.~\cite{Decca:2009fg}, who were considering gravitational
and Yukawa type forces, but we see it applies to any central force.

For example, the above, exact formula for weakly-coupled semitransparent
surfaces says that the force on such a sphere, of radius $a$, 
the center of which is a distance $Z$ above a semitransparent plane is
\be
F_T=-\frac{\partial E_T}{\partial Z}=-\frac{\lambda_1\lambda_2 aT}{8}
\int_{2\pi T(Z-a)}^{2\pi T(Z+a)} \frac{du}u\coth u.
\ee
The zero-temperature limit of this is
\be
F=-\frac{\lambda_1\lambda_2}{8\pi}\frac{a^2}{Z^2-a^2},
\ee
which may be alternatively derived from the zero-temperature
energy
\be
E=-\frac{\lambda_1\lambda_2 a^2}{16\pi}\int_{-1}^1\frac{d\cos\theta}{Z+a
\cos\theta}=-\frac{\lambda_1\lambda_2 a}{16\pi}\ln\frac{Z+a}{Z-a},
\ee
again, the exact PFA result.

\section{Interaction between two semitransparent spheres at nonzero
temperature}\label{spheres}
Consider now two spheres, of radius $a$ and $b$, respectively, with a
distance between their centers $R>a+b$. In terms of local coordinates
with origins at the centers of the two spheres, the semitransparent
potentials are
\be
V_1=\lambda_1\delta(r-a),\quad V_2=\lambda_2\delta(r'-b),
\ee
and let us further suppose that $\mathbf{R}$ lies along the $z$ axis
of both coordinate systems.  Then the squared distance between points
on the spheres is
\be
|\mathbf{r-r'}|^2=R^2+a^2+b^2-2ab\cos\gamma-2R(a\cos\theta-b\cos\theta'),
\ee
in terms of polar angles in the two spheres, where the cosine of the angle
between the two radial vectors locating the points is
\be
\cos\gamma=\cos\theta\cos\theta'+\sin\theta\sin\theta'\cos(\phi-\phi').
\ee
We insert this into the expression for the energy (\ref{ET}), obtaining
\be
E=-\frac{\lambda_1\lambda_2 T}{32\pi^2}a^2b^2\int d\Omega \,d\Omega'
\frac{\coth 2\pi T|\mathbf{r-r'}|}{|\mathbf{r-r'}|^2}.
\ee

It seems difficult to proceed in general, but we can work out a low
temperature expansion using
\be
\coth y= \sum_{n=0}^\infty \frac{2^{2n}B_{2n}}{(2n)!}y^{2n-1}
=\frac1y+\frac13 y-\frac1{45}y^3+\dots,\quad (|y|<\pi)
\ee
which will give rise to an expansion of the form
\be
E_T=E_0+T^2 E_2+T^4 E_4+\dots.
\ee
The zero temperature result was worked out, by inspection, in 
Ref.~\cite{Milton:2007wz}:
\be
E_0=-\frac{\lambda_1\lambda_2 ab}{16\pi R}\ln\frac{1-(a-b)^2/R^2}{1+(a+b)^2
/R^2}.\label{e0}
\ee
The $T^2$ term is trivial because it is evaluated by Newton's theorem
that a Coulomb potential exterior to a spherically symmetric 
charge distribution is as though the
charge were concentrated at the center:
\be
E_2=-\frac{\lambda_1\lambda_2\pi }{3}\frac{a^2b^2}R.\ee
The $T^{2n}$ term, $n>1$ however, is slightly nontrivial:
\be
E_{2n}=-\frac{\lambda_1\lambda_2}{64\pi^3}a^2b^2\frac{(4\pi)^{2n}B_{2n}}{(2n)!}
\int d\Omega\,d\Omega'|\mathbf{r-r'}|^{2n-3}.
\ee
 We may evaluate the integrals by expanding
in powers of $\hat a=a/R$ and $\hat b=b/R$:
\begin{subequations}
\bea
\int d\Omega\,d\Omega'|\mathbf{r-r'}|&=&(4\pi)^2 R\left[1+\frac13({\hat a}^2+
{\hat b}^2)\right]\\
\int d\Omega\,d\Omega'|\mathbf{r-r'}|^3&=&(4\pi)^2 R^3\left[1+2({\hat a}^2+
{\hat b}^2)+\frac15{\hat a}^4+\frac23{\hat a}^2{\hat b}^2
+\frac15{\hat b}^4\right]\\
\int d\Omega\,d\Omega'|\mathbf{r-r'}|^5&=&(4\pi)^2 R^5\bigg[1+5({\hat a}^2+
{\hat b}^2)+3{\hat a}^4+10{\hat a}^2{\hat b}^2
+3{\hat b}^4\nn\\
&&\quad\mbox{}+\frac17{\hat a}^6+{\hat a}^2{\hat b}^2({\hat a}^2+{\hat b}^2)
+\frac17{\hat b}^6\bigg],
\eea
\end{subequations}
and so on.  The reason these are polynomials is evident when one considers
the multipole expansion of the Coulomb potential---See, for example, Chap.~22
of Ref.~\cite{embook}.  For general formulas for such moments, see the
Appendix.

By computing further terms in the sequence of polynomials, we
are able to recognize the pattern:
\be
\frac1{(4\pi)^2 R^{2n+1}}\int d\Omega\,d\Omega' |\mathbf{r-r'}|^{2n+1}
=\sum_{p=0}^{n+1}\sum_{q=0}^p A(n,p,q) {\hat a}^{2(p-q)}{\hat b}^{2q},
\ee
where
\be
A(n,p,q)=\frac{(2n+2)!}{(2n-2p+2)!(2p-2q+1)!(2q+1)!}.
\ee
When this is inserted into the low temperature expansion, we can remarkably
sum the series:
\bea
E_T&=&-\frac{\lambda_1\lambda_2}{16 \pi}\frac{ab}R
\bigg\{\ln\frac{1-(a-b)^2/R^2}{1-(a+b)^2/R^2}\nn\\
&&\quad\mbox{}+f(2\pi T(R+a+b))+f(2\pi T(R-a-b))-f(2\pi T(R-a+b))
-f(2\pi T(R+a-b))\bigg\},\label{genform}
\eea
where $f$ is 
\be
f(y)=\sum_{n=1}^\infty \frac{2^{2n}B_{2n}}{2n(2n-1)(2n)!}y^{2n},
\label{ltexp}
\ee
which is obtained from the second antiderivative of the hyperbolic
cotangent:
\be
y\frac{d^2}{dy^2}f(y)=\coth y-\frac1y,\quad f(0)=f'(0)=0.\label{diffeq}
\ee
Although the power series expansion (\ref{ltexp}) is valid only for
sufficiently low temperatures $2T(R+a+b)<1$, the solution of the
differential equation is valid for all values of $T$.

For sufficiently high temperatures we can replace the hyperbolic
cotangent in the differential equation by 1, and then 
\be
f(y)\sim y\ln y+\ln y+A y+B,\quad y\gg1,
\ee
where $A$ and $B$ are integration constants that do not contribute to
Eq.~(\ref{genform}).  When this asymptotic solution is inserted into
Eq.~(\ref{genform}) the zero temperature logarithm cancels out, and we
are left with
\be
E_T\sim
-\frac{\lambda_1\lambda_2 a b}8 T\left[\ln\frac{R^2-(a+b)^2}{R^2-(a-b)^2}
+\frac{a}R\ln\frac{(R+b)^2-a^2}{(R-b)^2-a^2}+\frac{b}R\ln\frac{(R+a)^2-b^2}
{(R-a)^2-b^2}\right],\quad T\to\infty.\label{hit}
\ee

This result may be derived directly from the high-temperature form
\be
E_T\sim -\frac{\lambda_1\lambda_2 T a^2b^2}{32\pi^2}\int d\Omega\,d\Omega'
\frac1{|\mathbf{r-r'}|^2},\quad T\to\infty.
\ee
This again may be worked out by expanding in powers of the radii of the
spheres.  Computing the first several terms reveals the pattern:
\be
\int d\Omega\,d\Omega' \frac1{|\mathbf{r-r'}|^2}=\frac{(4\pi)^2}{R^2}
\sum_{n=0}^\infty \frac1{(n+1)(2n+1)}
\sum_{m=0}^n\frac12{2n+2\choose 2m+1} {\hat a}^{2(n-m)}
{\hat b}^{2m}.\label{dsum}
\ee
This sum is almost identical to that found for spheres at zero temperature,
as seen in Eq.~(6.15) of Ref.~\cite{Milton:2007wz}, which led to 
Eq.~(\ref{e0}), except for the appearance of $1/(2n+1)$ here.
Therefore, the former series must be
obtained from  the present series by differentiation. Denoting the
double sum in Eq.~(\ref{dsum}) by $S$, it must be true that
\be
R^2\frac{\partial}{\partial R}\frac{S}{R}=
\frac{R^2}{4 a b}\ln\left(\frac{1-(a+b)^2/R^2}{1-(a-b)^2/R^2}\right),
\ee
where $S$ is $R^2/4ab$ times the square-bracketed quantity in Eq.~(\ref{hit}).
This equality is, in fact, easily verified.  
See the Appendix for the generalization of this result.

We compare the general form [obtained by numerically integrating
Eq.~(\ref{diffeq})] and the high-temperature limiting form (\ref{hit}) in 
Fig.~\ref{fig1}.
\begin{figure}[tb]
  \begin{center}
  \includegraphics[width=3in]{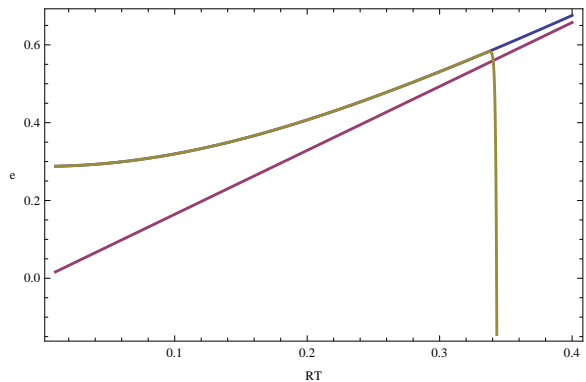} 
  \caption{Comparison between the general and high temperature forms
of the energy, as a function of $RT$. Energies are shown for $a=b=R/4$.
The high temperature result is linear in $T$. Also shown is the power
series expansion [Eq.~(\ref{ltexp}) truncated at 200 terms],
which diverges in this case at $RT=1/3$.  Plotted is 
$e=-16\pi R E/(\lambda_1\lambda_2 a^2)$.}
  \label{fig1}
  \end{center}
\end{figure}

\section{conclusions} We have shown that exact results may be found
in weak coupling for the quantum vacuum forces between nontrivial bodies
not only at zero temperature, but at finite temperature.  We have shown
that the exact form of the proximity force approximation holds exactly
for all temperatures for the force between an infinite plane surface and
an arbitrarily curved one.  We have also computed the force between
two semitransparent spheres at arbitrary temperatures,
and obtain remarkably simple, closed-form expressions.  The
PFA equivalence evidently will hold for tenuous dielectric bodies in
electromagnetism, and closed-form
finite temperature results may be easily obtained
between dielectric bodies as well.

\begin{acknowledgments}
K.A.M. first met Tom Erber at UCLA during one of Tom's 
many visits to Schwinger's
group there in the 1970s.  Thus it is a great pleasure to dedicate this
paper to him. 
We thank the US National Science Foundation (Grant No.\ PHY-0554926) and the
US Department of Energy (Grants Nos.\ DE-FG02-04ER41305
and DE-FG02-04ER-46140) for partially funding this research. 
We thank Nima Pourtolami and Elom Abalo for collaborative assistance.
\end{acknowledgments}

\appendix
\section{Mean powers of distances between points on spheres}
In Sec.~\ref{spheres} we used exact evaluations of mean distances, defined
by
\be
\int d\Omega\,d\Omega' |\mathbf{r-r'}|^p=(4\pi)^2 R^p P_p({\hat a},{\hat b}),
\ee
where $R$ is the distance between the centers of the two nonoverlapping
spheres, of radii $a$ and $b$, respectively.  Here $\hat a=a/R$ and
$\hat b=b/R$, and $P_p({\hat a},{\hat b})$ can in general be represented
by the infinite series
\be
P_p(\hat a,\hat b)=\sum_{n=0}^\infty \frac{2}{(2n+2)!}\frac{\Gamma(2n-p-1)}
{\Gamma(-p-1)}Q_n(\hat a,\hat b).
\ee
Here the homogeneous polynomials $Q_n$ are
\begin{subequations}
\bea
Q_0&=&1,\\
Q_1&=&2({\hat a}^2+{\hat b}^2),\\
Q_2&=&3{\hat a}^4+10{\hat a}^2{\hat b}^2+3{\hat b}^4,\\
Q_3&=&4{\hat a}^6+28{\hat a}^4 b^2+28{\hat a}^2{\hat b}^4+4{\hat b}^6,
\eea
\end{subequations}
or in general,
\be
Q_n=\frac12\sum_{m=0}^n{2n+2\choose 2m+1}{\hat a}^{2(n-m)}{\hat b}^{2m}.
\ee

We can easily see the following recursion relation holds:
\be
P_{p-1}({\hat a},{\hat b})=\frac{R^{-p}}{1+p}\frac\partial{\partial R}
R^{1+p}P_p(\hat a,\hat b),\label{rr}
\ee
since $Q_n$ is homogeneous in $R$ of degree $-2n$.

For $p$ a non-negative integer, $P_p$ is a polynomial of degree $2\lceil p/2\rceil $,
and we can immediately find
\be
P_p(\hat a,\hat b)=\frac1{4{\hat a}{\hat b}}\frac1{(p+2)(p+3)}
\left[(1+\hat a+\hat b)^{p+3}+(1-\hat a-\hat b)^{p+3}
-(1-\hat a+\hat b)^{p+3}-(1+\hat a-\hat b)^{p+3}\right],\quad p=0,1,2,\dots.
\ee
For $p$ a negative integer, we have
\begin{subequations}
\bea
P_{-1}&=&1,\\
P_{-2}&=&\frac1{4{\hat a}{\hat b}}\left[\ln\frac{1-({\hat a}+{\hat b})^2}
{1-({\hat a}-{\hat b})^2}+\hat a\ln\frac{(1+\hat b)^2-{\hat a}^2}
{(1-\hat b)^2-{\hat a}^2}+\hat b\ln\frac{(1+\hat a)^2-{\hat b}^2}
{(1-\hat a)^2-{\hat b}^2}\right]
,\\
P_{-3}&=&-\frac1{4{\hat a}{\hat b}}\ln\frac{1-({\hat a}+{\hat b})^2}
{1-({\hat a}-{\hat b})^2},\\
P_{-4}&=&\frac1{[1-({\hat a}+{\hat b})^2][1-({\hat a}-{\hat b})^2]},
\eea
\end{subequations}
and further expressions can be obtained by use of Eq.~(\ref{rr}).

\end{document}